\documentstyle[11pt,a4,cite,psfig]{article}
\newcommand{\lapprox}{%
\mathrel{%
\setbox0=\hbox{$<$}
\raise0.6ex\copy0\kern-\wd0
\lower0.65ex\hbox{$\sim$}
}}
\newcommand{\gapprox}{%
\mathrel{%
\setbox0=\hbox{$>$}
\raise0.6ex\copy0\kern-\wd0
\lower0.65ex\hbox{$\sim$}
}}

\newcommand{\be}{\begin{equation}}
\newcommand{\ee}{\end{equation}}
\newcommand{\bea}{\begin{eqnarray}}
\newcommand{\eea}{\end{eqnarray}}
\newcommand{\noi}{\noindent}

\newcommand{\ra}{\rightarrow}

\newcommand{\lesssim}{ {\
\lower-1.2pt\vbox{\hbox{\rlap{$<$}\lower5pt\vbox{\hbox{$\sim$}}}}\ }  }
\newcommand{\gtrsim}{ {\
\lower-1.2pt\vbox{\hbox{\rlap{$>$}\lower5pt\vbox{\hbox{$\sim$}}}}\ }  }

\newcommand{\cO}{{\cal O}}

\newcommand{\Imm}{\mbox{\rm Im}}

\newcommand{\MeV}{\mbox{\rm MeV}}
\newcommand{\GeV}{\mbox{\rm GeV}}

\newcommand{\annd}{\mbox{\rm and}}
\newcommand{\foor}{\mbox{\rm for}}

\newcommand{\EM}{\mbox{\rm {\scriptsize EM}}}

\begin{document}

\begin{titlepage}

\vspace*{2cm}

\begin{center} 

{\Large \bf Patterns of Spontaneous Chiral Symmetry Breaking\\
\vspace*{0.3cm}
in the Large--$N_c$ Limit of QCD--Like Theories  }\\[1.5cm] 

{\large {\bf Marc
Knecht} and  {\bf Eduardo de Rafael}}\\[1cm]
Centre  de Physique Th\'eorique\\ CNRS-Luminy, Case 907\\
F-13288 Marseille Cedex 9, France\\

\vspace*{1.0cm}

{\bf Abstract}

\end{center}

It is shown that in vector--like gauge theories, such as QCD, and in the limit of a
large number of colours $N_c$, the ordering pattern of narrow vector
and axial--vector states at low energies is correlated with the size of the
possible local order parameters of chiral symmetry breaking. This has
implications on the underlying dynamics of spontaneous chiral symmetry breaking
in QCD, and also on possible values of the parameter $S$ for oblique
electroweak
corrections, the analogue of the QCD low energy constant $L_{10}$.

\vspace{5cm}

\noi

\indent

\noi December 1997

\noi
CPT-97/P.3574

\bigskip

\noindent anonymous ftp or gopher: cpt.univ-mrs.fr


\end{titlepage}

\begin{titlepage}

$\ $

\end{titlepage}

{\large{\bf 1 Introduction}}
\vspace*{3 mm}

\noi
Sometime ago, Coleman and Witten~\cite{CW80} showed that if QCD with
the number of colours $N_c=3$ confines, and if this property
persists in the large--$N_c$ limit~\cite{'tH74,RV77,W79} then, in this limit,
the chiral
$U(n_f)\times U(n_f)$ invariance of the Lagrangian  with
$n_f$ flavours of massless quarks is spontaneously
broken down to the diagonal $U(n_f)$ subgroup. Their proof assumes in
particular that ``the
breakdown of chiral symmetry is characterized by a nonzero value of some order
parameter which is bilinear in the quark fields and which transforms according
 to the representation $(n_f,\bar{n}_f)+(\bar{n}_f,n_f)$ of the chiral group''. A few years later, the argument was substantiated by a result due to Vafa and
Witten~\cite{VaWi}. These authors showed that in vector--like theories, the
generators associated with the diagonal subgroup always leave the vacuum
invariant in the theory with massive quarks. Combined with 't Hooft's
anomaly matching conditions \cite{tHo}, this leads, in the large--$N_c$
limit, and without further assumptions besides confinement~\cite{CW80}, to
the expected pattern of spontaneous chiral symmetry breaking (S$\chi$SB),
at least for the massless theory defined by the zero mass limit of the
massive one\footnote{An additional requirement in \cite{VaWi} is that all
vacuum angles vanish.}. As further emphasized by the authors of
ref.~\cite{CW80}, these results do not ``give any insight at all into the
mechanism of symmetry breakdown''.
The
purpose of this note is to show how, in the same large--$N_c$ limit, the
ordering
pattern of possible vector and axial--vector hadronic states in the physical
spectrum is correlated to the size of the possible order parameters.

We shall consider a correlation function which is particularly sensitive to
properties of S$\chi$SB, namely the
two--point function
\be
\Pi_{LR}^{\mu\nu}(q)=2i\int d^4 x\,e^{iq\cdot x}\langle 0\mid
T\left(L^{\mu}(x)R^{\nu}(0)^{\dagger}\right)\mid 0\rangle \,,
\ee with currents
\be L^{\mu}=\bar{d}(x)\gamma^{\mu}\frac{1}{2}(1-\gamma_{5})u(x)
\qquad \annd \qquad
R^{\mu}=\bar{d}(x)\gamma^{\mu}\frac{1}{2}(1+\gamma_{5})u(x)\,.
\ee
In the chiral limit where the light quark masses are set to zero, this
two--point function only depends on one invariant function ($Q^2=-q^{2}\ge 0$
for
$q^2$ spacelike)
\be
\Pi_{LR}^{\mu\nu}(q)=(q^{\mu}q^{\nu}-g^{\mu\nu}q^2)\Pi_{LR}(Q^2)\,.
\ee
We shall be concerned with properties of the self--energy function
$\Pi_{LR}(Q^2)$ in the limit of vanishing light quark masses and infinite
number of colours.
Some of these
properties are well known. First,
$\Pi_{LR}(Q^2)$ vanishes order by order in perturbation theory and is an
order parameter of the spontaneous breakdown of chiral symmetry for all
values of the momentum transfer.
It also governs the electromagnetic $\pi^{+}-\pi^{0}$ mass
difference~\cite{Lowetal67}
\be\label{eq:piem}
m_{\pi^+}^{2}\vert_{\EM}=\frac{\alpha}{\pi}\,\frac{-3}{8f_{\pi}^2}\,
\int_0^\infty
dQ^2\,Q^2\Pi_{LR}(Q^2)\ .
\ee
This integral converges in the ultraviolet region
because, as shown by Shifman, Vainshtein and Zakharov~\cite{SVZ79} using
Wilson's \cite{Wil} operator product expansion (OPE),
\be\label{eq:OPE}
\lim_{Q^2\ra\infty}\Pi_{LR}(Q^2)=\frac{1}{Q^6}
\left[-8\pi^2\left(\frac{\alpha_s}{\pi}+\cO(\alpha_s^2)\right)
\langle\bar{\psi}\psi\rangle^2
\right]+\cO\left(\frac{1}{Q^8}\right)\,.
\ee
Witten~\cite{W83, CLT95} has furthermore shown that
\be\label{eq:witten}
-Q^2\Pi_{LR}(Q^2)\ge 0 \qquad\foor\;  \quad 0\le Q^2\le\infty\,,
\ee
which in particular ensures the positivity of the integral in
eq.~(\ref{eq:piem}) and thus the stability of the QCD vacuum with respect
to  small perturbations induced by electromagnetic interactions.

The low $Q^2$ behaviour of this self--energy function~\cite{BR91} is
governed by chiral perturbation theory
\be
-Q^2\Pi_{LR}(Q^2)=f_{\pi}^2+4L_{10}Q^2+\cO(Q^4)\,,
\ee
where $L_{10}$ is
one of the Gasser--Leutwyler constants~\cite{GL85} of the
$\cO(p^4)$ low energy effective chiral Lagrangian, i.e. the Lagrangian
formulated in terms of Goldstone degrees of freedom and external local
sources only.

In the large--$N_c$ limit, the spectral function associated to
$\Pi_{LR}(Q^2)$ consists of the difference of
an infinite number of narrow vector states and an infinite number of narrow
axial--vector states, together with the Goldstone pole of the pion:
\be
\frac{1}{\pi}\Imm\Pi_{LR}(t) =\sum_{V}f_{V}^2 M_{V}^2\delta(t-M_{V}^2)
-\sum_{A}f_{A}^2 M_{A}^2\delta(t-M_{A}^2)-f_{\pi}^2\delta(t)\,.
\ee
Since $\Pi_{LR}(Q^2)$ obeys an unsubtracted dispersion relation, it
follows that
\be\label{eq:LRN1}
-Q^2\Pi_{LR}(Q^2)=f_{\pi}^2+\sum_{A}f_{A}^2 M_{A}^2\frac{Q^2}{M_{A}^2+Q^2}
-\sum_{V}f_{V}^2 M_{V}^2\frac{Q^2}{M_{V}^2+Q^2}\,.
\ee

\vspace*{7mm}
{\large{\bf 2 Weinberg Sum Rules}}
\vspace*{3 mm}

\noi
In the chiral limit, the first Weinberg sum rule~\cite{We67}
\be
\int_{0}^{\infty} dt \,\Imm\Pi_{LR}(t)=0\,,
\ee
implies that
\be\label{eq:1wsr}
\sum_{V}f_{V}^2 M_{V}^2-\sum_{A}f_{A}^2 M_{A}^2=f_{\pi}^2\,;
\ee
the second Weinberg sum rule~\cite{We67}
\be
\int_{0}^{\infty} dt\,t \,\Imm\Pi_{LR}(t)=0\,,
\ee
furthermore implies that
\be\label{eq:2wsr}
\sum_{V}f_{V}^2 M_{V}^4-\sum_{A}f_{A}^2 M_{A}^4=0\,.
\ee
In QCD~\footnote{For a discussion of the Weinberg sum rules in QCD in the
presence of explicit chiral symmetry breaking see ref.~\cite{FNR79}},
eq.~(\ref{eq:1wsr}) follows from the fact that there is no local order
parameter of dimension $d=2$ and eq.~(\ref{eq:2wsr}) from the absence of a
local order parameter of dimension $d=4$. The first possibly non trivial
contribution comes from local order parameters of dimension $d=6$, as shown
in eq.~(\ref{eq:OPE}). With these constraints incorporated
into eq.~(\ref{eq:LRN1}), the
self--energy function in the large--$N_c$ limit
becomes
\be\label{eq:LRN2}
-Q^2\Pi_{LR}(Q^2)=\sum_{A}f_{A}^2 M_{A}^6\frac{1}{Q^2(M_{A}^2+Q^2)}
-\sum_{V}f_{V}^2 M_{V}^6\frac{1}{Q^2(M_{V}^2+Q^2)}\,.
\ee

There are further properties of this self--energy function, perhaps not so
well known,  which we now wish to emphasize:

\begin{itemize}
\item
The expansion of the hadronic self--energy function in the r.h.s. of
eq.~(\ref{eq:LRN2}) in inverse
powers of
$Q^2$:
\be\label{eq:LRhadronic}
\Pi_{LR}(Q^2)=\frac{1}{Q^6}(\sum_{V} f_{V}^2 M_{V}^6-\sum_{A} f_{A}^2 M_{A}^6) -
\frac{1}{Q^8}(\sum_{V} f_{V}^2 M_{V}^8-\sum_{A} f_{A}^2 M_{A}^8) +
\cdots\,,
\ee
has to match the short--distance OPE evaluated in the QCD large--$N_c$
limit~\footnote{Notice that the OPE of this self--energy function is free
from renormalon ambiguities, because in perturbation theory it is protected by
chiral symmetry, which ensures that $\Pi_{LR}(Q^2)=0$ order by order in powers
of $\alpha_{s}$.}.
This means e.g., that in this limit the leading $d=6$ order parameter in
eq.~(\ref{eq:OPE}) is given by
\be\label{eq:phi3}
-8\pi^2\left(\frac{\alpha_s}{\pi}+\cO(\alpha_s^2)\right)
\langle\bar{\psi}\psi\rangle^2=\sum_{V} f_{V}^2
M_{V}^6-\sum_{A} f_{A}^2 M_{A}^6\equiv\phi^{(6)}\,.
\ee

\item
Witten's inequality in eq.~(\ref{eq:witten}) implies in particular
that
\be
\lim_{Q^2 \ra\infty}-Q^2 \Pi_{LR}(Q^2)\ge 0\,.
\ee
From this inequality and eq.~(\ref{eq:LRhadronic}) there follows the
remarkable result that
$\phi^{(6)}$ in (\ref{eq:phi3}) must be negative (or zero), a fact which, as
eq.~(\ref{eq:OPE}) shows, is  indeed confirmed  by the explicit calculation
to lowest
non--trivial order~\cite{SVZ79}.

\item
Inverse moments of the spectral function with the pion pole
removed, i.e. integrals like
\be
\int_{0}^{\infty}dt\frac{1}{t^p}\left(\frac{1}{\pi}\Imm\Pi_{LR}(t)
+f_{\pi}^2\delta(t)\right)=
\int_{0}^{\infty}dt\frac{1}{t^p}\left(\frac{1}{\pi}\Imm\Pi_{V}(t)
-\frac{1}{\pi}
\Imm\Pi_{A}(t)\right)\,,
\ee
with $p=1,2,3,\cdots$, correspond to non--local order parameters which
govern the couplings of
local operators of higher and higher dimensions in the low energy effective
chiral Lagrangian. For example, the first inverse moment is related to the
$\cO(p^4)$ coupling constant $L_{10}$ as follows~\cite{GL84}
\be
-4L_{10}=\int_{0}^{\infty}\frac{dt}{t}\,\left(\frac{1}{\pi}\Imm\Pi_{V}(t)-
\frac{1}{\pi}
\Imm\Pi_{A}(t)\right)=
\sum_{V}f_{V}^2-\sum_{A}f_{A}^2\,.
\ee

\item
One of the parameters which characterize possible deviations from the
Standard Model predictions in the electroweak sector is the so called
$S$--parameter~\cite{PT90}. In the unitary gauge, $S$ measures the strength
of an anomalous $W^{(3)}_{\mu\nu}B^{\mu\nu}$ coupling. This is the
$SU(2)_{L}\times SU(2)_{R}$ coupling analogous to the term proportional to
$L_{10}$ in QCD. It has been argued that if the underlying theory of
electroweak breaking is a vector--like gauge theory of the QCD--type, the sign of the
$L_{10}$--like coupling must be negative, precisely as observed in QCD. In the
electroweak sector, this fact is infirmed  by experimental observation and
constitutes at present a serious phenomenological obstacle to
technicolour--like formulations of electroweak symmetry breaking. As we shall
see, the {\it sign} of the
$L_{10}$ coupling depends on the relative size of the local order parameters
versus
$f_{\pi}^2$.

\end{itemize}

\vspace*{7mm} {\large{\bf 3 Solving Systems of Weinberg Sum Rule Equations}}
\vspace*{3 mm}

\noi
Given an arbitrary, but finite, number $n$ of narrow states of spin 1
$\{R_1,R_2,\cdots R_n\}$,
we want to know which configurations of vector and axial vector narrow states
are compatible with large--$N_c$.  We find it convenient to use the following
notation: the squared masses and squared
couplings will be denoted
\be
X_{i}\equiv M_{R_{i}}^2 \qquad\annd\qquad
x_{i}\equiv \pm f_{R_{i}}^2\,,\quad i=1,2,3,\cdots n\,,
\ee
with the $+f_{R_{i}}^2$ choice if $R_{i}$ is a vector state, and the
$-f_{R_{i}}^2$ choice if $R_{i}$ is an axial--vector state. The states
are ordered according to increasing values of their masses,
\be
X_1\le X_2\le\cdots X_n\,.
\ee
Cast into this notation, the first and second Weinberg sum
rules result then in the following system of two
equations:
\be
\left\{\begin{array}{ccc}
x_{1}X_{1}  +x_{2}X_{2}  +x_{3}X_{3}+  \cdots +x_{n}X_{n}  & = &  f_{\pi}^2 \\
x_{1}X_{1}^2+x_{2}X_{2}^2+x_{3}X_{3}^2+\cdots +x_{n}X_{n}^2  & = & 0\,.
\end{array}\right.
\ee
Implicit here is the assumption that the sum of the infinite number of narrow
vector states and the sum of the infinite number of narrow axial--vector
states with masses higher than the mass of the last
 narrow state $R_n$ explicitly considered are already dual to their respective
perturbative QCD continuum, so that their contributions cancel in the spectral
function $\Imm\Pi_{LR}(t)$, and hence in the Weinberg sum rules.

Further linear equations in the couplings $x_{i}$, involving higher powers
of the
squared masses
$X_{i}$, result from identifying inverse powers of ${Q^{2}}$  in the OPE of the
self--energy function $\Pi_{LR}$,
\be
\Pi_{LR}(Q^2)=\sum_{p\ge 6}\frac{1}{Q^p}\phi^{(p)}\,,
\ee
with the corresponding expansion of its hadronic counterpart in
eq.~(\ref{eq:LRhadronic}). We can thus write a system of $n$ linear
equations for the $n$ couplings $x_{i}$:
\be\label{eq:fullweqs}
\left\{\begin{array}{ccc} x_{1}X_{1}  +x_{2}X_{2}  +x_{3}X_{3}+  \cdots
+x_{n}X_{n}  & = &  f_{\pi}^2 \\ x_{1}X_{1}^2+x_{2}X_{2}^2+x_{3}X_{3}^2+\cdots
+x_{n}X_{n}^2  & = & 0 \\
 x_{1}X_{1}^3+x_{2}X_{2}^3+x_{3}X_{3}^3+\cdots +x_{n}X_{n}^3  & = &
\phi^{(6)} \\
\cdots &  & \\
 x_{1}X_{1}^n+x_{2}X_{2}^n+x_{3}X_{3}^n+\cdots +x_{n}X_{n}^n  & = &
\phi^{(2n)} \,.
\end{array}\right.
\ee
The discriminant of this system is a Vandermonde determinant
\bea
\Delta(X_1,X_2,X_3,\cdots X_n) & = & X_1 X_2 X_3\cdots X_n \times
\prod_{1\leq i< j \leq n}\,(X_j-X_i)\,,
\eea
and the system has a solution. There are some interesting generic properties
which emerge from this system of equations which we next discuss.

\begin{itemize}

\item
The coupling constant $f_{\pi}^2$ is a non local order parameter of spontaneous
chiral symmetry breaking. Assuming that $f_{\pi}^2\not=0$, but setting the
chiral condensates
$\phi^{(6)}=\phi^{(8)}=\cdots=\phi^{(2n)}=0$, the system in
eq.~(\ref{eq:fullweqs}) has a solution with alternate signs for the
couplings $x_{1}\,,x_{2}\,\cdots\,x_{n}$, and with $x_1>0$. Therefore the
ordering pattern
of states in that case has to be $V_{1}-A_{2}-V_{3}-A_{4}-\cdots$, i.e.
alternating vector and axial--vector states.

\item
Once the system in eq.~(\ref{eq:fullweqs}) has been solved under the assumption
that $\phi^{(6)}=\phi^{(8)}=\cdots=\phi^{(2n)}=0$, the sizes of the
chiral condensates of dimension $d\ge 2n+2$ are then fixed. In particular,
\be
\phi^{(2n+2)}=(-1)^{n+1}\,f_{\pi}^2\,X_{1}X_{2}X_{3}\cdots\,X_{n}\,,
\ee
which cannot vanish. We then conclude that for QCD in the large--$N_c$
limit {\it
spontaneous chiral symmetry breaking \`{a} la Nambu--Goldstone with
$f_{\pi}^2\not=0$ necessarily implies the existence of non--zero local order
parameters which transform according to the representation
$(n_f,\bar{n}_f)+(\bar{n}_f,n_f)$ of the chiral group.} This is in a way the
converse of the Coleman--Witten theorem~\cite{CW80} stated in the
Introduction.

\item
The sign of the lowest dimensional non--zero chiral condensate is fixed, and
agrees with Witten's inequality.

\end{itemize}

\noi
Let us next discuss the simplest systems of Weinberg sum rule equations
with $n=2$ and $n=3$ in further detail.


\vspace*{5mm}
{\bf 3a The Two--State Case}
\vspace*{2mm}

\noi
This is the case originally considered by Weinberg~\cite{We67},
and also in resonance dominance estimates of the low
energy constants of the effective chiral Lagrangian~\cite{EGPR89,EGLPR89,DRV89}.
With the notation defined previously, the simplest case of
two states
$\{R_{1},R_{2}\}$
has the solutions:
\be\label{eq:x1y1}
 x_{1}=f_{\pi}^2\frac{1}{X_1}\frac{X_2}{X_{2}-X_{1}},
\quad , \qquad x_{2}=-f_{\pi}^2\frac{1}{X_2}\frac{X_1}{X_{2}-X_{1}}\,.
\ee
This results in
\be
-4L_{10}\equiv x_{1}+x_{2}=f_{\pi}^2\left(\frac{1}{X_{1}}+
\frac{1}{X_{2}}\right)\,,
\ee
and therefore the coupling $L_{10}$ in the two--state case has to be
negative~\cite{Pe80,Pr81}. Several comments are in order.

\begin{itemize}

\item
We see that the two--state case is indeed a particular case of what has
been stated above. It is the fact that $f_{\pi}^2\not=0$ that allows for a
non--trivial solution. Since
$X_{1}< X_{2}$, we find that
$x_{2}< 0$. Therefore the second resonance $R_{2}$  must be an axial--vector
state, and the first one a vector state.

\item
Once the two--state system is solved, all the higher dimension condensates
which break spontaneously the chiral symmetry are fully fixed by the masses
$X_{1}$ and
$X_{2}$ only; in particular the $d=6$ condensate which appears in the OPE, see
eq.~(\ref{eq:OPE}), is now fixed to be
\be\label{eq:phi6}
\phi^{(6)}\equiv x_{1}X_{1}^3 +x_{2}X_{2}^3 =-f_{\pi}^2 X_{1}X_{2}\,,
\ee
and is indeed negative.

\item
Notice that within the simple two--state pattern $V_{1}-A_{2}$, the
possibility that
$\langle\bar{\psi}\psi\rangle=0$, which would imply $\phi^{(6)}=0$,
contradicts eq.~(\ref{eq:phi6}). Therefore, we conclude that the extreme
version of the so called {\it generalized}
$\chi$PT proposed by J.~Stern {\it et al.}~\cite{FSSKM}, where
$\langle\bar{\psi}\psi\rangle=0$, and which we shall denote as
G$\chi$PT$_{0}$ in
what follows, is incompatible with the simplest realization of an hadronic low
energy spectrum with only one $V$--state and only one $A$--state.

\end{itemize}

\vspace*{5mm}
{\bf 3b The Three--State Case}
\vspace*{2mm}

\noi
Let us next discuss the case of three narrow states $\{R_1,R_2,R_3\}$ with
increasing ordering in masses: $ X_{1}\le X_{2} \le X_{3}$.
The corresponding system of Weinberg equations is now
\be\label{eq:3weqs}
\left\{\begin{array}{ccc}
x_{1}X_{1}+x_{2}X_{2}+x_{3}X_{3} & = & f_{\pi}^2  \\
x_{1}X_{1}^2+x_{2}X_{2}^2+x_{3}X_{3}^2 & = & 0  \\
x_{1}X_{1}^3+x_{2}X_{2}^3+x_{3}X_{3}^3 & = & \phi^{(6)}\,,
\end{array}\right.
\ee
where we know that in the third equation $\phi^{(6)}\le 0$, but we leave its
size as a free parameter. The solution of this system of equations  is
\be\label{eq:x1}
 x_{1} =
\frac{f_{\pi}^2 X_{2}X_{3}+\phi^{(6)}}{X_{1}}
\ \frac{1}{(X_{2}-X_{1})(X_{3}-X_{1})}\,,
\ee
\be\label{eq:x2}
x_{2} =
\frac{f_{\pi}^2 X_{3}X_{1}+\phi^{(6)}}{X_{2}}
\ \frac{-1}{(X_{2}-X_{1})(X_{3}-X_{2})}\,,
\ee
\be\label{eq:x3}
x_{3} =
\frac{f_{\pi}^2 X_{1}X_{2}+\phi^{(6)}}{X_{3}}
\ \frac{1}{(X_{3}-X_{1})(X_{3}-X_{2})}\,;
\ee
and the sum of these solutions gives the following prediction for $L_{10}$:
\be
-4L_{10}=x_1 +x_2
+x_3 =f_{\pi}^2\left(\frac{1}{X_{1}}+\frac{1}{X_{2}}+\frac{1}{X_{3}}
\right) +\frac{\phi^{(6)}}{X_{1}X_{2}X_{3}}\,.
\ee
It follows from these equations that the pattern of $V$--states versus
$A$--states is now governed by the size of the $d=6$ order parameter
$\phi^{(6)}$. We observe the following generic facts:

\begin{itemize}

\item
In the
large--$N_c$ limit, $\langle\bar{\psi}\psi\rangle =0$ not only entails
$\phi^{(6)}=0$, but also $\phi^{(8)}=0$, since the $d=8$ chiral condensate
becomes  also proportional to
$\langle\bar{\psi}\psi\rangle$. This contradicts the result
\be\label{eq:con8}
\phi^{(8)}=f_{\pi}^2 X_{1}X_{2}X_{3}\,,
\ee
which follows from eqs.~(\ref{eq:x1}) to (\ref{eq:x3}). Therefore, we conclude
that G$\chi$PT$_{0}$ is incompatible both with a hadronic low energy spectrum of
one vector state and one axial--vector state as well as with a low energy
spectrum of two vector states and one axial state.

\item
In QCD with
$\langle\bar{\psi}\psi\rangle\not=0$,
$\phi^{(6)}$ must be negative. From the solutions for the couplings
$x_{1}\,,x_{2}\,,x_{3}$ above it follows that the ordering pattern will be
$V_{1}-A_{2}-V_{3}$ provided
$f_{\pi}^2 X_{1}X_{2}+\phi^{(6)}>0$ .

\item
When
$f_{\pi}^2 X_{1}X_{2}+\phi^{(6)}=0$ ,
$x_{3}=0$ and the highest vector state decouples. We are then back to the
two--state case
$V_{1}-A_{2}$  already discussed in the previous subsection
{\bf 3a}. When
$f_{\pi}^2 X_{1}X_{2}+\phi^{(6)}<0$ ,
the spectrum pattern becomes $V_{1}-A_{2}-A_{3}$, until $\phi^{(6)}$ is
sufficiently negative,
$\phi^{(6)}=-f_{\pi}^2 X_{1}X_{3}$,
where the lowest axial state decouples.
 For
$f_{\pi}^2 X_{1}X_{3}+\phi^{(6)}<0$ ,
the pattern becomes $V_{1}-V_{2}-A_{3}$ and remains so until
$f_{\pi}^2 X_{2}X_{3}+\phi^{(6)}=0$ ,
where $x_{1}=0$ and the first vector state decouples.
Finally, for
$f_{\pi}^2 X_{2}X_{3}+\phi^{(6)}<0$ ,
the pattern becomes $A_{1}-V_{2}-A_{3}$, and it remains like that as
$\phi^{(6)}$ becomes more and more negative.

\item
In all cases where
$f_{\pi}^2 X_{2}X_{3}+\phi^{(6)}\ge 0$ ,
we find that $-4L_{10}>0$. However, as $\phi^{(6)}$ becomes more and more
negative, there is a critical value
\be
\hat{\phi}^{(6)}=-f_{\pi}^2\left(X_{2}X_{3}+X_{1}X_{3}+X_{1}X_{2}\right)\,,
\ee
at which $-4L_{10}=0$. Beyond that critical value $-4L_{10}<0$. The
pattern for positive values of $L_{10}$ has to be $A_{1}-V_{2}-A_{3}$.

\end{itemize}


\centerline{\psfig{figure=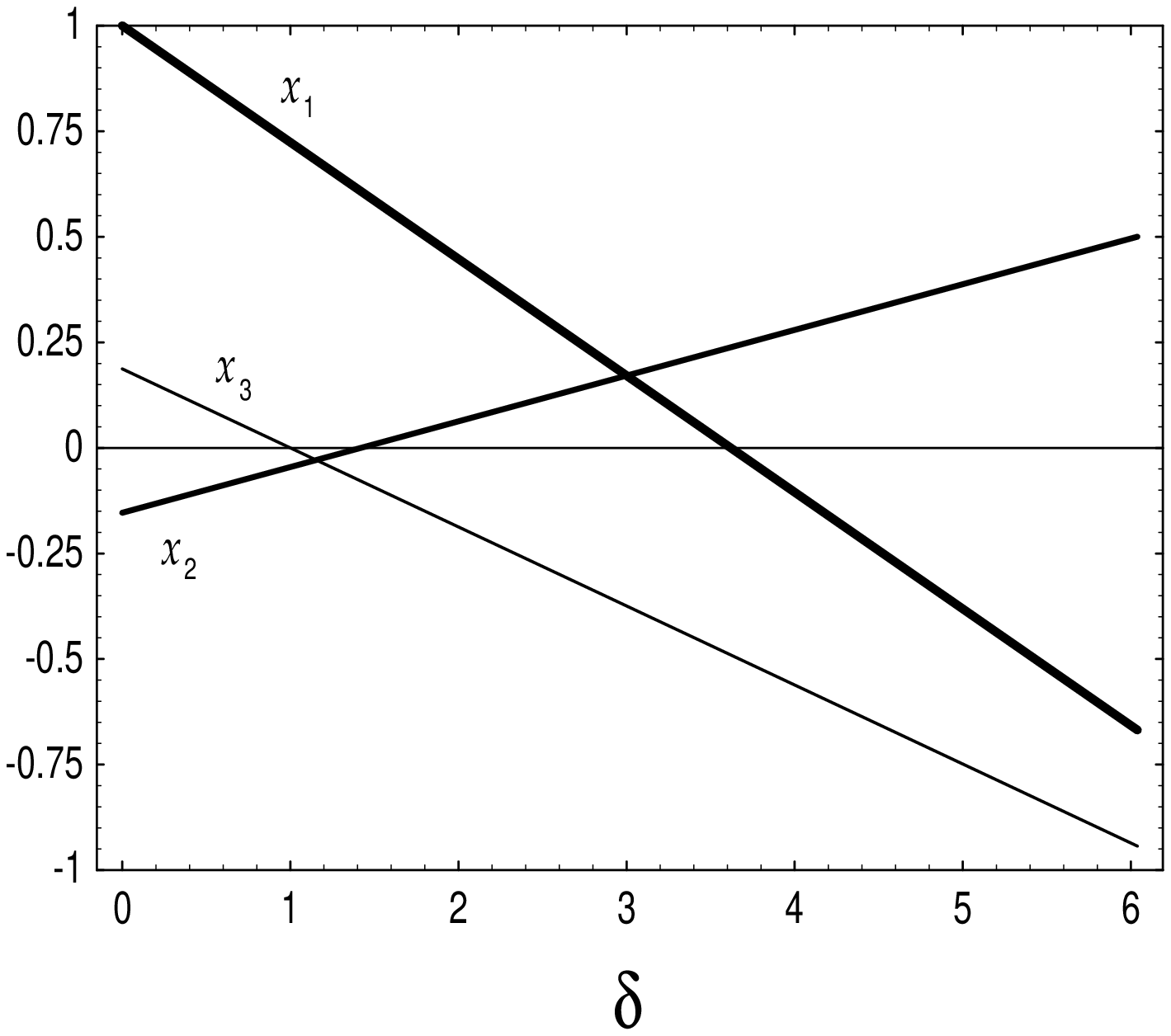,height=12cm}}
{\bf Fig.~1} {\it The three normalized couplings
$x_{1}\,, x_{2}\,, x_{3}$  versus the parameter $\delta$ of
eq.~(\ref{eq:delta}).}

\vspace*{5mm}

The evolution of patterns as a function of $\phi^{(6)}$ is illustrated in
Fig.~1 for a choice of the mass squared values $X_{1,2,3}$ as given by the
central values of the masses of the observed lowest hadronic vector and
axial--vector resonances,
$X_{1}=0.6\,\GeV^2\,,X_{2}=1.5\,\GeV^2\,,X_{3}=2.1\,\GeV^2$. In this
figure we plot the predicted couplings
$x_{1}$,$x_{2}$, and $x_{3}$, normalized to their overall maximum,  as functions
of the dimensionless parameter $\delta$ defined as follows
\be\label{eq:delta}
\phi^{(6)}=-f_{\pi}^2 X_{1}X_{2}\delta\,.
\ee
Positive values of $x_{i}$ correspond to vector states and negative values of
$x_{i}$ to axial--vector states. For
$\delta=0.5$, for instance, the pattern is $V_{1}-A_{2}-V_{3}$ in that order of
increasing masses, while for $\delta=5$ the pattern is $A_{1}-V_{2}-A_{3}$ also
in that order of increasing masses. With this choice of masses, the
constant $L_{10}$ becomes positive for $\delta\gapprox 6$.

\vspace*{7mm}
{\large{\bf 4 Phenomenological Implications}}
\vspace*{3 mm}

\noi
There are several phenomenological implications which follow from the analyses
reported in the previous sections, and which we now wish to discuss.

\vspace*{5mm}

{\bf 4a QCD Low Energy Phenomenology and the Two--State Case}
\vspace*{2mm}

\noi
The phenomenological ansatz that the hadronic spectrum in the vector and
axial-vector channels can be well
approximated by a single dominant low energy narrow state plus a
perturbative QCD
continuum with an onset adjusted by global duality arguments~\cite{BLR85}
has been shown to be rather successful in many instances. It is at the basis of
the QCD sum rule approach pioneered by the ITEP group~\cite{SVZ79}. Moreover,
as already mentioned, it is the underlying assumption in the resonance dominance
estimates of the low energy constants of the effective chiral
Lagrangian~\cite{EGPR89,EGLPR89,DRV89}, estimates which are rather
successful. The phenomenological successes of models,
like the extended
Nambu--Jona-Lasinio model~\footnote{See e.g. ref.\cite{Bij} for a recent review
article where many other references can be found.} are also correlated with
this ansatz.  It can be shown~\cite{PPR97} that the large--$N_c$ limit
 of QCD and the OPE provide a logical support to these
successes. One can see {\it a posteriori} that the
approximation works because of the relatively small size of the perturbative
$\Lambda_{QCD}$--scale. The onset of the dual continuum, in the vector
and axial--vector channels, happens at values where perturbative QCD can
already be
trusted. The analysis of section {\bf 3a} above shows that in the large--$N_c$ limit the two--state pattern is
necessarily correlated with the usual picture, where spontaneous breakdown
of chiral symmetry is triggered through the formation of a strong
$\langle\bar{\psi}\psi\rangle$ condensate. Indeed, from eqs.~(\ref{eq:phi3}) and
(\ref{eq:phi6}) it follows that in this case
\be
8\pi^2\,\frac{\alpha_s}{\pi}\langle\bar{\psi}\psi\rangle^2\,\simeq
f_{\pi}^2\,X_{1}X_{2}\,.
\ee
Using the experimental central values of $f_{\pi}$, $M_{\rho}$ and $M_{A1}$,
and with $\alpha_s$ evaluated at the corresponding scale $s_0$ of global duality
($s_0\simeq 1.6\,\GeV^2$), this equation predicts a quark condensate
$\langle\bar{\psi}\psi\rangle (1\,{\rm GeV})\simeq -(270\,\MeV)^3$ which,
within the expected
errors of the approximations involved, is compatible with the values which
can be obtained
from other phenomenological estimates~\cite{BPR95}.

\vspace*{5mm}

{\bf 4b The Generalized Chiral Perturbation Theory Alternative}
\vspace*{2mm}

\noi
The formation of a large $\langle\bar{\psi}\psi\rangle$ condensate is
usually accepted as being the actual mechanism of S$\chi$SB. This
assumption is dropped in G$\chi$PT~\cite{FSSKM}, which considers the
possibility that the
$\langle\bar{\psi}\psi\rangle$ condensate may not be the
dominant order parameter of S$\chi$SB. It turns out that it is
not easy to
distinguish phenomenologically this possibility from the standard one with
the present
precision of low energy measurements. As we have demonstrated, in the combined
large--$N_c$ and chiral limit, the extreme version of G$\chi$PT,
where the quark condensate would vanish exactly
$\langle\bar{\psi}\psi\rangle=0$, requires a rather complex structure of
the large--$N_c$
spectrum before asymptotic freedom sets in.
The minimum structure compatible with a vanishing quark condensate, but with a
non--zero mixed condensate:
\be
\langle\bar{\psi}\psi\rangle=0,\qquad
\langle\bar{\psi}\sigma_{\mu\nu}G^{\mu\nu}\psi\rangle\not=0\,,
\ee
is two vector states and two axial--vector states
with the ordering pattern: $V_{1}-A_{2}-V_{3}-A_{4}$.
It is the mixed
condensate $\langle\bar{\psi}\sigma_{\mu\nu}G^{\mu\nu}\psi\rangle$
which, even in the absence of a quark condensate, can give a term with
dimension
$d=10$ in the OPE.

It is interesting to ask how the $V_{1}-A_{2}-V_{3}-A_{4}$
pattern compares with what is presently known from experiment~\cite{PDB96}. The
answer is shown in Fig.~2 where we display
the central mass values of the observed vector and axial--vector resonances in
the isospin $I=1$ (Fig.~2a) and $I=1/2$ (Fig.~2b) channels.
In the $I=1/2$ channel, the situation is not far from reproducing the
required pattern. On the other hand, there is no established second
axial--vector state in the $I=1$ channel.
Hadronic $\tau$ decays are at present the best place to observe a
possible second $I=1$ axial--vector state, although the phase space
limitations are severe.
From a phenomenological point of view and with
$\Lambda_{QCD}\lapprox 300\,\MeV$  (in the $\overline{MS}$--scheme),  there is no
compulsory need to fill the hadronic low energy spectrum with more than a
prominent narrow state in the vector and axial--vector channels in order to
obtain
duality with perturbative QCD. In that sense, the option of S$\chi$SB with
a vanishing quark condensate, although not  
rigorously
ruled out, seems unnatural. On the other hand, the
present analysis cannot exclude large $1/N_c$--corrections in the
factorization of the vacuum expectation values of four--quark
operators, and
hence the possibility that in the real world the quark condensate might
be smaller, say
$\langle\bar{\psi}\psi\rangle\sim -f_{\pi}^3$, than what is found in present
estimates and than what is usually believed.

\vspace*{5mm}

{\bf 4c Implications for Models of Electroweak Symmetry Breaking}
\vspace*{2mm}

\noi
Perhaps the most interesting observation which emerges from the
analyses reported in the previous sections is the possible existence of low
energy particle spectra in vector--like  gauge theories with a rather different
structure than the one observed in the QCD hadronic spectrum. The simple
three--state case discussed in section 3b shows already an explicit example
of how the
ordering pattern in the spectrum and the value of the electroweak $S$
parameter are explicitly correlated to the size of the $d=6$
condensate. From this example, it seems plausible to consider
technicolour--like models with a sufficiently large number of fermions so that
the $\beta$--function enhances the value of the running coupling at the
``pertinent'' duality scale and makes the $d=6$ condensate strongly negative,
and hence $S$ negative as well.

We hope that these observations may open new paths in model building
of electroweak symmetry breaking.

\vspace*{7mm} {\large{\bf Acknowledgments}}
\vspace*{3 mm}

\noi
We are grateful to J. Bijnens, S. Peris and J. Stern for
helpful discussions at various stages of this work.

\newpage

\vspace*{3 mm}

\centerline{\psfig{figure=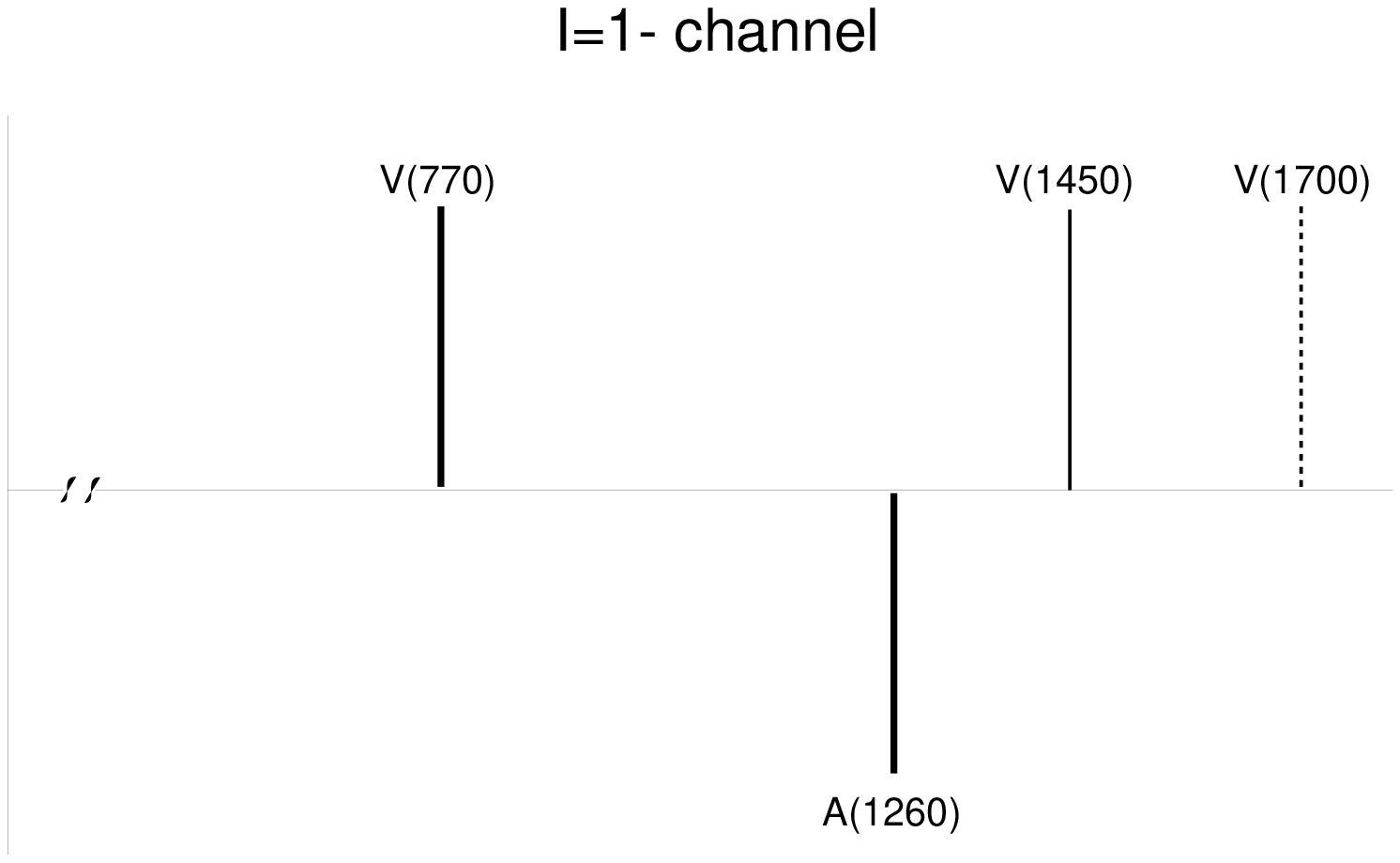,height=8cm}}
{\bf Fig.~2a} {\it Low lying vector and
axial--vector resonances in the isospin $I=1$ channel.}

\vspace*{7mm}

\centerline{\psfig{figure=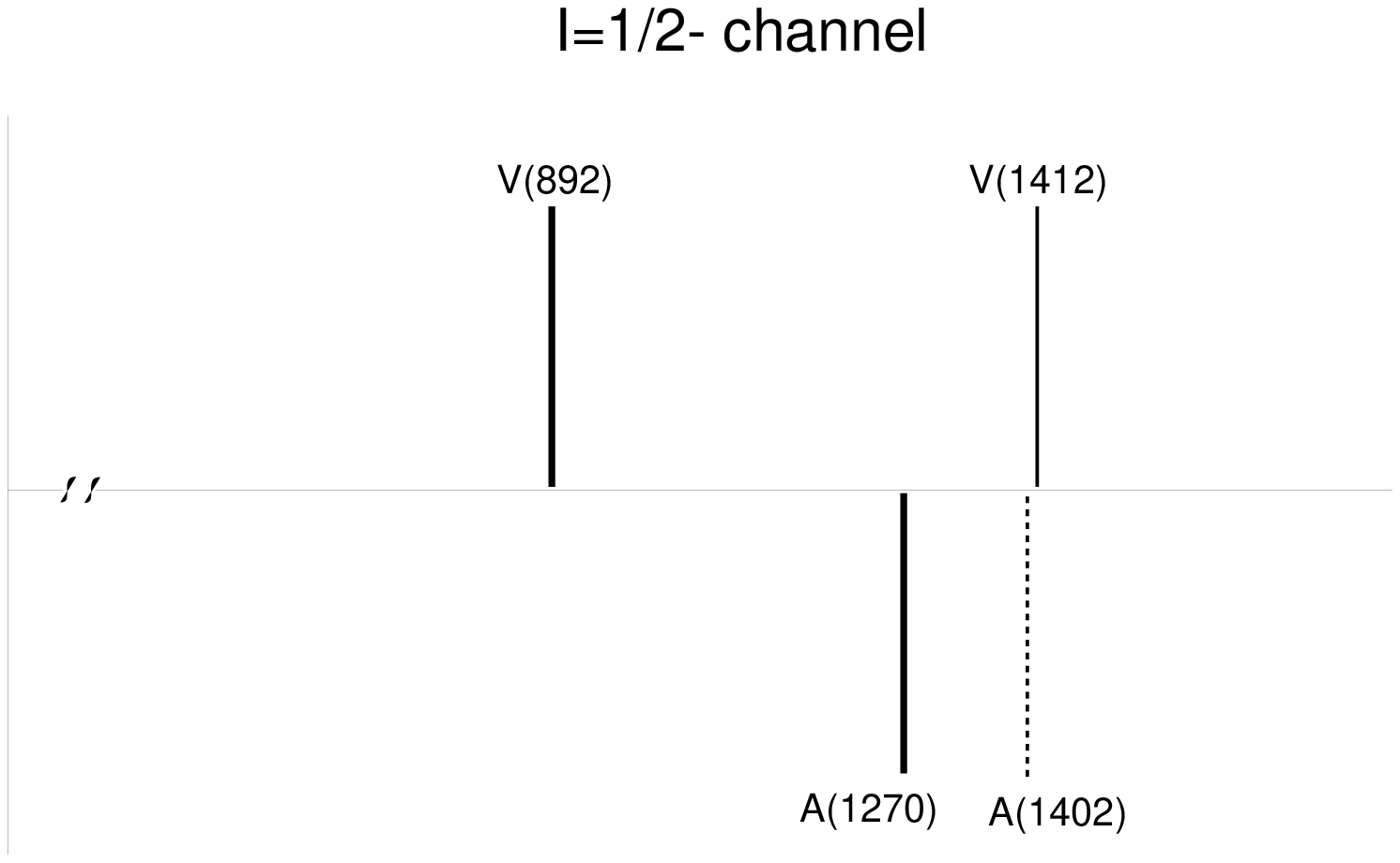,height=8cm}}
{\bf Fig.~2b} {\it Low lying vector and
axial--vector resonances in the isospin $I=1/2$ channel.}



\newpage




\end{document}